\documentclass[twocolumn,superscriptaddress,groupedaddress,amsmath,amssymb,aps,prl,floatfix]{revtex4-2}

\usepackage{comment}
\usepackage{graphicx}
\usepackage{dcolumn}
\usepackage{bm}

\usepackage[utf8]{inputenc}
\usepackage[T1]{fontenc}

\usepackage{braket}
\usepackage{siunitx}
\usepackage{xcolor}

\usepackage[normalem]{ulem}

\def\Rb87{^{87}\mathrm{Rb}}                             
\def\He4{^{4}\mathrm{He}}     

\def\ex{\mathbf{e}_x}  
\def\ey{\mathbf{e}_y}  
\def\ez{\mathbf{e}_z}

\begin{document}

\title{Kolmogorov scaling in turbulent 2D Bose-Einstein condensates}

\author{M.~Zhao}
\author{J.~Tao}
\author{I.~B.~Spielman}
\email{ian.spielman@nist.gov}
\affiliation{Joint Quantum Institute, University of Maryland and National Institute of Standards and Technology, College Park, Maryland 20742, USA }

\date{\today} 

\begin{abstract}
We investigated turbulence in 2D atomic Bose-Einstein condensates (BECs) using a minimally destructive, impurity injection technique analogous to particle image velocimetry in conventional fluids.
Our approach transfers small regions of the BEC into a different hyperfine state, and tracks their displacement ultimately yielding the velocity field. 
This allows us to quantify turbulence in the same way as is conventional in fluid dynamics in terms of velocity-velocity correlation functions, called velocity structure functions, that obey Kolmogorov scaling laws.
Furthermore the velocity increments show a clear fat-tail non-Gaussian distribution that results from intermittency corrections to the initial ``K41'' Kolmogorov theory.
Our observations are fully consistent with the later ``KO62'' description.
These results are validated by a 2D dissipative Gross-Pitaevskii simulation. 
\end{abstract}

\maketitle
Turbulence is a fundamental phenomenon encountered in a wide range of fluids at all scales: from classical systems such as oceans and atmospheres~\cite{gargett1989ocean,wyngaard1992atmospheric}; confined and solar plasmas~\cite{zhou2004colloquium,loureiro2017role}; and the self-gravitating media of the large-scale universe~\cite{shandarin1989large} to quantum fluids such as neutron stars~\cite{greenstein1970superfluid}, superfluid $^4$He ~\cite{vinen2002quantum} and atomic Bose-Einstein condensates (BECs) ~\cite{henn2009emergence,navon2016emergence}.
In all of these cases, turbulence is characterized by complex patterns of fluid motion spanning a wide range of length scales.
While the understanding of classical turbulence has matured in the past century~\cite{sreenivasan1999fluid}, that of quantum systems has many open questions~\cite{madeira2020quantum}.
For example, in BECs, does there exist a range of length scales---the inertial scale---in which kinetic energy cascades from large to small scale (``direct,'' as in 3D classical fluids) or from small scale to large (``inverse,'' as for 2D classical fluids) in accordance with a Kolmogorov scaling law?
Although Kolmogorov scaling was predicted only for incompressible fluids it has been observed in virtually all turbulent fluids~\cite{sreenivasan1999fluid}. 
Kolmogorov scaling is generally quantified in terms of velocity structure functions (VSFs) that require knowledge of the fluid's velocity field, which is difficult to measure in quantum gas experiments.
In this work we: present a particle image velocimetry (PIV) technique~\cite{taylor1938spectrum,adrian1985pulsed,bradley2006decay,guo2010visualization} employing spinor impurities as tracer particles; obtain VSFs in 2D turbulent atomic BECs; and experimentally observe Kolmogorov scaling.

Existing experimental evidence for turbulence in atomic BECs relies on time of flight (TOF) measurements are either dominated by interaction driven expansion~\cite{henn2009emergence}, or yield momentum distributions~\cite{thompson2013evidence,navon2016emergence}.
Such observations have no clear connection to the VSFs $S_p(l)$ which describe various order-$p$ moments of the distribution of velocity increments 
\begin{align}
\delta {\bf v}(\mathbf{x}, {\bf l}) &= \mathbf{v}(\mathbf{x}+\mathbf{l})-\mathbf{v}(\mathbf{x})
 \label{eq:StructureFunction}
\end{align}
as a function of displacement ${\bf l}$.
Without access to VSFs, turbulence in atomic gases lacks a direct point of comparison to other fluids.

Unlike classical fluid flow, superfluid flow is irrotational (with vorticity confined to the cores of quantized vortices, where the superfluid density is zero) with a velocity field governed by the phase of the superfluid order parameter $\phi$ via ${\bf v} = \hbar{\boldsymbol \nabla}\phi/m$.
Despite this, it is generally believed that superfluid turbulence obeys the same $S_p(l)\propto l^{(p/3)}$ scaling as classical fluids, described by the initial K41 Kolmogorov theory~\cite{kolmogorov1991local, kolmogorov1941degeneration, kolmogorov1991dissipation}; in the case of $\He4$ this has been experimentally verified~\cite{nore1997kolmogorov, stalp1999decay} for $p\leq 3$.
The more complete KO62 theory~\cite{kolmogorov1962refinement,obukhov1962some} adds an intermittency correction that becomes important for large $p$ and also predicts that the ensemble probability density function of velocity increments (PDF) is non-Gaussian, with ``fat-tails.''
Power-law scaling behavior and energy cascade have been observed in the momentum distribution of homogeneously trapped BECs undergoing relaxation~\cite{navon2016emergence}; the observed exponent departed from the $-5/3$ prediction of K41 theory for energy cascade [related to the Fourier transform of the $S_2(l)$ structure function], and was instead accurately interpreted using a wave turbulence model.

\begin{figure*}[t]
\includegraphics{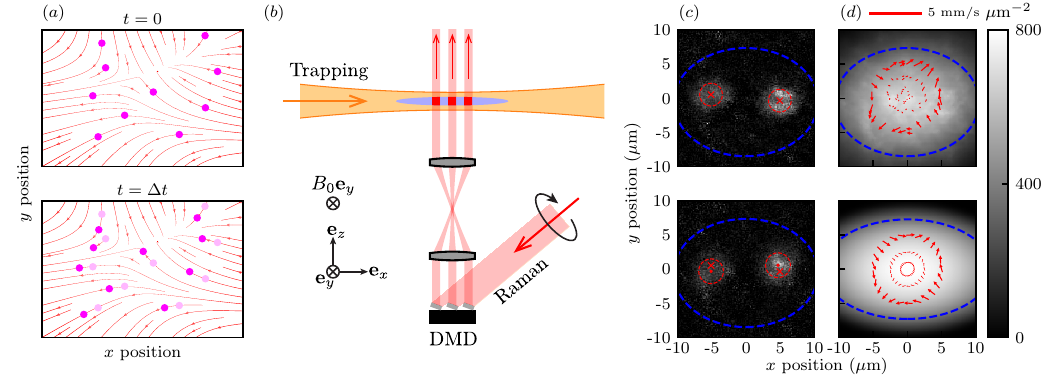}
\caption{
Concept. (a) Top: representative velocity field with tracer particles (bright pink). 
Bottom: at $t=\Delta t$, tracer particles move from their initial position (dark pink).   
(b) Schematic of spatially-resolved Raman apparatus used to create tracers.
(c) PTAI imaged tracers before (top) and after (bottom) evolution, with initial and final positions shown by crosses and circles respectively. 
Red circles mark the $1$-$\sigma$ widths of the tracers.  
(d) Velocity fields in a rotating harmonic trap (red arrows) along with atomic density in grey-scale with blue ellipses placed at $0.75\times$ of the Thomas-Fermi radius.
Top and bottom panels display experimental data and GPE simulations respectively.
}
\label{fig:1}
\end{figure*}

Our cold-atom PIV technique allows us to directly measure the velocity field and thereby both $S_p(l)$ and the underlying PDF.
As illustrated in Fig.~\ref{fig:1}(a), we prepare an initial velocity distribution, then create localized ``tracer particles'' consisting of atoms in a different hyperfine state using a spatially-resolved technique, and, after a $\Delta t$ delay, measure the tracers' displacement.
This then directly leads to the local fluid velocity.

{\it Experimental method}---We used $\Rb87$ BECs with $N\approx2\times10^5$ atoms in the $\ket{F=1, m_F=1}$ hyperfine ground state with strong vertical confinement [trap frequency $\omega_z/(2\pi) = 220\ \mathrm{Hz}$] provided by a $1064\ \mathrm{nm}$ laser with an elliptical cross-section, traveling along $\ex$.
Additionally, a digital micromirror device (DMD) patterned a $638\ \mathrm{nm}$ multimode laser traveling along $-\ez$ to provide dynamical ($\approx3\ {\rm kHz}$ update rate) in-plane potentials $V({\bf r}, t)$.
An $\approx 0.14\ {\rm mT}$ bias magnetic along $\ey$ created a $\Delta f = 1\ \mathrm{MHz}$ Zeeman splitting between consecutive $m_F$ states.

Figure~\ref{fig:1}(b) schematically shows the spatially-resolved Raman setup used to create localized tracer particles. 
A circularly polarized bichromatic $\approx790\ \mathrm{nm}$ laser beam traveling along $\ez$ with frequencies spaced by $\Delta f$ drove $m_F$-changing Raman transitions with a $50\ \mathrm{kHz}$ Rabi frequency.
The beam was patterned by a second DMD, enabling the placement of arbitrary patterns of tracer atoms in $\ket{F=1, m_F=0}$.
The inter- and intra-state interaction strengths for these hyperfine states differ only at the $0.5\ \%$ level~\cite{Widera2006}; as a result, tracers co-move with the underlying fluid~\footnote{Even for completely symmetric interactions the overall fluid flow can be slightly modified.
Physically this results from a contribution to the kinetic energy inherent in the tracer wavefunction, and is negligible in our experiments.}. 
Tracer atoms were selectively measured using partial transfer absorption imaging (PTAI), in which $\approx6.8\ \mathrm{GHz}$ microwaves transferred the tracers to $\ket{F=2, m_F=0}$ where they were detected using resonant absorption imaging.
Our imaging system had a nominal $1\ \mu{\rm m}$ resolution, allowing us create and then detect tracers with $1/e$ radius down to $1.6\ \mu{\rm m}$.

In our experimental sequence we first initialized the velocity field of interest and then created a set of $N$ tracers, at positions $\mathbf{r}^j_0$ in the $\ex$-$\ey$ plane using an $\approx \pi/2$ Raman pulse, with $j=1\ldots N$ [these positions were directly verified by PTAI measurement, as shown for $N=2$ in Fig.~\ref{fig:1}(c-top)].
After a $\Delta t$ evolution time, we imaged the tracers to obtain the final positions $\mathbf{r}^j$ [Fig.~\ref{fig:1}(c-bottom)].
The velocity at each $\mathbf{r}_0^j$ was taken as the first order finite-difference ${\bf v}^j = (\mathbf{r}^j - \mathbf{r}_0^j)/\Delta t$. 

{\it Validation in a rotating BEC}---
Before applying PIV to turbulent systems, we validated the method with harmonically trapped BECs rotating with angular frequency $\Omega$ about $\ez$.
The confining DMD generated a rotating in-plane harmonic potential with $(\omega_x, \omega_y) = 2\pi \times (40, 50)\ \mathrm{Hz}$.
For slowly rotating systems, such that no vortices are present, the superfluid velocity is expected to exhibit an irrotational pattern
$\mathbf{v} = a(y\mathbf{e}_x + x\mathbf{e}_y)$ with $a\propto\Omega$ for small $\Omega$~\cite{Pitaevsiii2003}.
At higher rotation frequencies, when $\Omega$ becomes comparable to the trap frequencies $\omega_{x,y}$, this becomes a metastable configuration with a range of possible instability conditions, the details of which must be obtained numerically~\cite{recati2001overcritical, sinha2001dynamic}.
In our case these conditions would limit the rotation frequency to $\Omega \lesssim 2\pi \times 40\ {\rm Hz}$, leading to typical speeds $|{\bf v}| \lesssim 0.25\ \mathrm{mm/s}$.
To obtain increased signal, we focused on overcritical systems with $\Omega = 2\pi\times50\ {\rm Hz}$, for which $|\mathbf{v}| \approx 0.7\ \mathrm{mm/s}$.

Experimentally we began with static systems, then linearly increased the angular frequency from zero to $\Omega$ in $15\ \mathrm{ms}$, held $\Omega$ constant for $2\ \mathrm{ms}$ (at which time the BEC rotated by an angle $\theta = \pi/2$) and then performed PIV.
In this demonstration we sampled the velocity field on three concentric circles (with radii of $1\ \mathrm{\mu m}$, $2.5\ \mathrm{\mu m}$, and $5\ \mathrm{\mu m}$) with an angular resolution of $\pi/12$, and used an evolution time $\Delta t = 1.5\ \mathrm{ms}$.
To increase the signal we used tracers with a $2.2\ \mathrm{\mu m}$ diameter and for the smallest circle the tracer size limited the number of tracers to $N=1$, otherwise we used $N=2$.
Figure~\ref{fig:1}(d-top) shows both the atomic density (grey scale) and associated velocity field (red) with the irrotational quadrupole pattern clearly visible.
The data is in good agreement with our Gross-Piteavskii equation (GPE) simulations in Fig.~\ref{fig:1}(d-bottom).
This validation of our PIV method also marks the first direct visualization of the irrotational flow pattern in a rotating atomic superfluid~\footnote{Going beyond the data presented here, we also validated the technique using dipole and scissors modes in non-rotating systems.}.

{\it Structure function}---Because Kolmogorov theory is valid for isotropic homogeneous systems, we turned our attention to near-ground state BECs with uniform atomic density.
We employed the confining DMD to create a time-independent 2D disk-shaped potential $V({\bf r}) = V_{\rm tr}  \Theta(|{\bf r}| - r_{\rm tr})$ with radius $r_{\rm tr} = 22\ \mathrm{\mu m}$ and depth $V_{\rm tr} \gg \mu$, where $\mu\approx h\times 550\ {\rm Hz}$ is the BEC's chemical potential~\footnote{We used the relation $\mu= 3 mc^2/2$, suitable for quasi-2D BECs with $\mu > \hbar\omega_z$, to obtain the chemical potential from the measured long wavelength speed of sound $c=1.3\ \mathrm{mm/s}$.}.

\begin{figure}[t]
\includegraphics{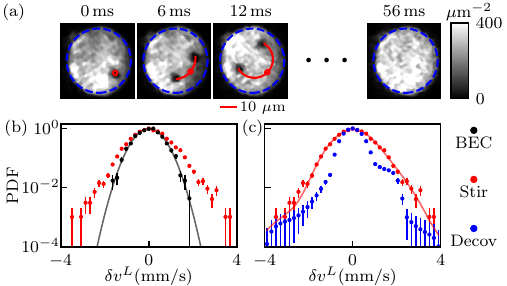}
\caption{Turbulence.
(a) Atomic density measured during the excitation process. The blue circle marks the edge of the trap and red arcs indicate the average path of the stirring rods.
(b,c) Histograms of $\delta v^{L}(l=10.6\ \mu m)$ with peak value scaled to $1$ and error bars derived from standard counting uncertainty.
(b) Raw data for initial BEC (black, along with Gaussian fit) and with stirring (red).  
(c) Deconvolved data (blue) and raw data (red, along with a recolvolved curve).
}
\label{fig:turbulence}
\end{figure}

We then initialized turbulence with a pair of counter-rotating stirring ``rods'' with $3.5\ \mathrm{\mu m}$ radii (also created by the confining DMD) that locally depleted the atomic density.
As shown in Fig.~\ref{fig:turbulence}(a), the initially overlapping rods followed nominally circular trajectories (red curves, with a  $25\ \mathrm{Hz}$ rotational frequency), the radius of which changed every $400\ \mu{\rm s}$ to a random value in the interval $[12\ \mathrm{\mu m}, 15\ \mathrm{\mu m}]$.
The stirring potential was applied for $16\ \mathrm{ms}$; the system was then allowed to relax for $40\ \mathrm{ms}$ prior to PIV measurement (with $\Delta t=0.3\,\mathrm{ms}$ evolution time). 

We used tracer patterns consisting of $N=4$ tracers arrayed in a square with three different side-lengths: $10.6\,\mathrm{\mu m}$, $11.4\,\mathrm{\mu m}$, and $12.6\,\mathrm{\mu m}$.
Together these patterns gave access to six tracer separations $l$ comprising the side as well as the diagonal lengths.
The measured tracer positions $\mathbf{r}^j$ were then identified by the center of mass $\overline{\rho^j\mathbf{r}}/\overline{\rho^j}$ of the transferred atoms.
To first order in $\Delta t$ the resulting velocity is the density-weighted (i.e., Favre-averaged~\cite{favre1965equations}) velocity $\widetilde{\mathbf{v}} =\overline{\rho \mathbf{v}}/\overline{\rho}$, used when applying Kolmogorov theory to compressible fluids~\cite{aluie2011compressible, aluie2013scale, wang2012scaling} (we omit the tilde in what follows).
Each measurement yielded 12 velocity increments $\delta {\bf v}(\mathbf{r}^j, {\bf l}^{ij})$, with ${\bf r}^j$ associated with each tracer and the difference vector ${\bf l}^{ij} = \mathbf{r}_0^i-\mathbf{r}_0^j$ to each of the remaining tracers.
Figure~\ref{fig:turbulence}(b) displays the resulting longitudinal PDFs both
with (red symbols) and without (black symbols) stirring, ${\rm PDF}_{\rm s}(\Delta v^{L})$ and ${\rm PDF}_{\rm ns}(\Delta v^{L})$ respectively. 
A ground state BEC's PDF should resemble a Kronecker-$\delta$ function centered at $0$; here the observed non-zero width $\sigma=0.55(1)\ \mathrm{mm/s}$ provides a measure of the instrumental noise and is well described by a Gaussian (black curve).
Although the distribution with stirring (red) is broadened and acquires a ``fat tail,'' any VSFs computed directly from these data will be significantly contaminated by instrumental noise. 

We therefore employed quadratic programming based deconvolution~\cite{yang2020density} to approximate the underlying ${\rm PDF_0}(\Delta v)$ from raw data [Fig.~\ref{fig:turbulence}(c)].
This process minimizes the L2 distance $|{\rm PDF}_{\rm s}(\Delta v) - ({\rm PDF}_{\rm ns} *{\rm PDF}_0 )(\Delta v)|^2$ between the measured distribution and the convolution (denoted by $*$) of the reconstruction with the instrument noise distribution, subject to the constraints that ${\rm PDF_0}(\Delta v)$ is: normalized, non-negative, and contains a single maximum. 
Because ``fancy'' analysis procedures may introduce unknown artifacts, in what follows we present data derived from the PDF both with and without deconvolution.

\begin{figure}[t]
\includegraphics{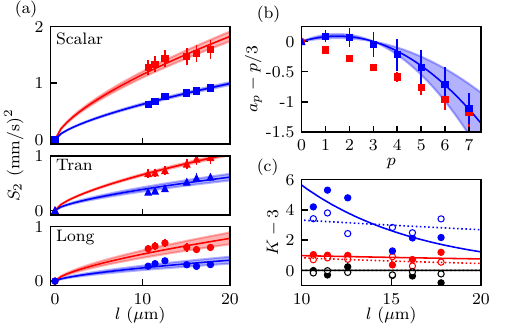}
\caption{ 
In all panels red and blue denote stirred data with and without deconvolution, respectively, while black indicates unstirred data.
(a) Measured $S_2(l)$.
Each point results the average of 44 experimental runs, and each run derived $S_2(l)$ from about 50 nominally identical experimental repetitions. 
The uncertainties are the $2$-$\sigma$ standard error of the mean for the set of 44 runs.
As described in the text, the curves are fits to the data plotted along with their $2$-$\sigma$ uncertainty band.
(b) Intermittency correction with error bars indicating $2$-$\sigma$ uncertainties.
The blue curve represents a fit to the KO62 model, with $2$-$\sigma$ confidence band.
(c) Excess Kurtosis of transverse (solid) and longitudinal (empty) distributions data each fit to a decaying exponential. 
}
\label{fig:3}
\end{figure}

Using PDFs such as these, we obtained the longitudinal $S^L_p(l) = \langle|\delta {\bf v}(\mathbf{x}, {\bf l}) \cdot {\bf e}_l|^p\rangle$, transverse  $S^T_p(l) = \langle|\delta {\bf v}(\mathbf{x}, {\bf l}) \cdot {\bf e}_\perp|^p\rangle$, and scalar $S^S_p(l) = \langle|\delta {\bf v}(\mathbf{x}, {\bf l})|^p\rangle$ VSFs~
\footnote{
The average $\left<\cdots\right>$ denotes the ensemble average over all positions ${\bf x}$ and displacement directions ${\bf e}_l$.
}.
All three 2nd order VSFs derived from this procedure are shown in Fig.~\ref{fig:3}(a) with and without deconvolution.
The primary impact of deconvolution on these data is to reduce the amplitude of the VSFs, as would be expected from the PDF's reduced width.
In both cases the data are compatible with the $S_2(l)=s_2 l^{2/3}$ power-law expected in the K41 assumption, with fits shown by the solid curves (coefficients shown in Table~\ref{tab:summary}).
In general, transverse VSFs are expected to be larger than their longitudinal counterparts; for homogenous and isotropic turbulence the second order structure functions have the exact relation $S^{T}_2(l) / S^{L}_2(l) = 4 / 3$, and indeed we find $s_2^{T} / s_2^{L} = 1.6(1)$ with deconvolution and $1.33(4)$ without.

\begin{table}[b!]
\begin{tabular}{c c c c c c} 
\hline\hline
    & $s_2^L$  & $ s_2^T$ & $s_2^S$ &  $\chi$ & \\
    & & $\times10^{-3}\ \mathrm{m^{4/3}/s^2}$ & & &  \\
\hline
Expt. Raw & 1.06(7)  & 1.41(2)  & 2.47(6) & N.A. & \\
Expt. Deconv. & 0.52(4) & 0.83(4) &  1.34(2) & 0.04(1) & \\
Numerics & 0.329(1) & 0.401(2) & 0.730(3) & 0.023(1) & \\
\hline\hline\\
\end{tabular}
\caption{\label{tab:summary} Fit parameters.}
\end{table}

{\it Intermittency}---Intermittency in turbulence can be quantified by corrections to K41's $l^{(p/3)}$ scaling law.
We directly obtain scaling exponents $a_p$ from power-law fits to the measured $p$-order scalar VSFs; the resulting differences $\varepsilon_p=a_p - p/3$ are plotted in Fig.~\ref{fig:3}(b) for stirred data with and without deconvolution (blue and red squares respectively).
All three cases show a deviation from $p/3$ scaling that grows with increasing $p$.
In detail, the deconvolved data is consistent with the $\varepsilon_p = -\chi p(p-3)$ prediction of KO62 theory (solid curve), with a system specific intermittency coefficient $\chi$ obtained by fitting the deconvolved data in Table~\ref{tab:summary}.

The deviation from K41 predictions indicates scale-dependent non-Gaussian behavior in the PDFs. 
Two widely separated tracers should have uncorrelated motion, resulting in Gaussian PDFs for the velocity increments. 
However, if correlations develop with decreasing tracer separation, the resulting PDFs can become non-Gaussian as we observe in Fig.~\ref{fig:turbulence}(b-c).
We describe the non-Gaussian behavior of these distributions in terms of the kurtosis $K$, which quantifies the relative weight of a distribution's tail with respect to its center; a Gaussian distribution has $K=3$.
Figure~\ref{fig:3}(c) plots the excess kurtosis $K-3$, computed from both the transverse (solid markers) and  longitudinal (empty markers) PDFs, as a function of tracer separation for unstirred data (black), raw data (red), and decolvolved data (blue); the exponentially falling curves serves as guides to the eye. 
As expected, without stirring the measured distributions are Gaussian ($K-3\approx0$) and acquire fat tails ($K-3 > 0$) in the turbulent case.
The transverse excess kurtosis rises with falling $l$, as would be expected when intermittency is important at smaller scales.
Unexpectedly the longitudinal $K-3$ is independent of $l$, and to gain further insight we turn to numerical simulations.

{\it Numerical simulation}---We conclude by comparing to numerical simulations of a dissipative Gross-Pitaevskii equation (dGPE) introduced in Ref.~\onlinecite{kobayashi2005kolmogorov} for the study of turbulent BECs.
The dGPE is given by
\begin{align}
i\hbar\partial_t \psi  &= e^{-i\kappa(\hat{\mathbf{k}})} \left(\frac{\hbar^2 \hat k^2}{2m} + V + g |\psi|^2 - \mu \right) \psi \label{eq:dGPE}
\end{align}
where $\mu$ is the chemical potential, and ${\kappa}(\mathbf{k}) = \kappa_0\Theta(|\mathbf{k}|-k_{\mathrm{cut}})$ introduces dissipation that damps excitations with wavelengths smaller than $2\pi/k_\mathrm{cut}$~\footnote{Our parameters relate to the notation in Ref.~\onlinecite{kobayashi2005kolmogorov} via $k_\mathrm{cut}=2\pi s/\xi$, where $\xi$ is the healing length and $s$ is a free tuning parameter.}.
In our experiment this is physically motivated by the evaporation process which constantly removes high energy excitations.  
Our $T \approx 20\ {\rm nK}$ temperature corresponds to a thermal phonon wavenumber $k_{\rm th} \approx 2\ \mu{\rm m}^{-1}$; we set $k_{\rm cut}=5\ \mu{\rm m}^{-1}$, at the boundary between the highly occupied condensate mode and the sparsely occupied thermal modes~\cite{Blakie2008}, and confirmed that the simulation results were unchanged by factors of 2 increase or decrease of $k_{\rm cut}$.
By contrast, the dissipation strength $\kappa_0$ was empirically set to $0.02$ to match that nominal scale of the experimental deconvolved $S_2$. 
 
\begin{figure}[t]
\includegraphics[scale=1]{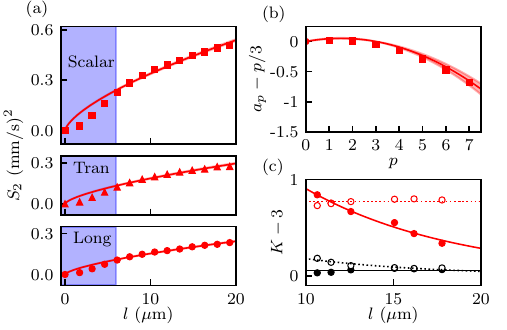}
\caption{
Numerical results.
In all panels, the colors, markers, fitting and uncertainty notations follow those in Fig.3.
(a) Numerical $S_2(l)$ obtained from the average of 40 numerical simulations.
In each run, $S_2(l)$ was calculated using a $6\ \mathrm{\mu m}$ region of interest for $\bar{v}$ and a $3.5\ \mathrm{\mu m}$ sampling distance for collecting velocity increments.
The blue region marks distances below the
$\approx 6\ \mathrm{\mu m}$ resolvable distance between tracers.
(b) Intermittency correction.
(c) Excess Kurtosis.
}
\label{fig:4}
\end{figure}

Each numerical experiment began with a steady state system evolving according to Eq.~\eqref{eq:dGPE} with an added stochastic noise term selected to give the observed $98\ \%$ condensate fraction.
The simulation then followed the experimental stirring / evolution protocol, and recorded the density weighted velocity averaged over the extent of the $\approx6\ \mu {\rm m}$ resolvable distance between tracers.

The numerical results in the remainder of Fig.~\ref{fig:4} parallels the experimental data in Fig.~\ref{fig:3}.
Panel (a) plots the second order VSFs where the blue shaded region deliniates the minimum resolvable distance between tracers.
The solid curves are fits to the $2/3$ scaling law outside of this regime with amplitudes shown in Table~\ref{tab:summary}. 
While the simulation parameter $\kappa_0$ was selected to match the nominal scale of these amplitudes with experiment, the close correspondence of their ratios---e.g.  $s^{T} / s^{L} = 1.22(1)$---as well as the overall scaling behavior are intrinsic outcomes of the simulation.
Figure~\ref{fig:4}b continues by showing the intermittency corrections $\varepsilon_p$ from scalar VSFs (red squares) are consistent with the KO62 theory (solid curve) with an intermittency exponent $\chi$ about half that of experiment (Table~\ref{tab:summary}).
Panel (c) shows that, as with experiment, $K-3$ increases from zero in the turbulent case, although with a significant reduction in overall magnitude.
As observed experimentally, the transverse excess kurtosis falls with increasing $l$, and the longitudinal $K-3$ remains independent of $l$.

Taken together these numerical results confirm the presence of vortex driven turbulence in this system, and provide near quantitative agreement with our experiment [this is quite surprising given the ad-hoc introduction of dissipation into Eq.~\eqref{eq:dGPE}].

{\it Discussion and outlook}---
Although our observations demonstrate Kolmogorov scaling behavior, our study leaves a range of open questions.
For example, Ref.~\cite{numasato2010direct} numerically showed a direct energy cascade is expected under conditions such as ours, rather than the inverse cascade usually associated with 2D systems.
In addition to energy and particle number, incompressible 2D fluids also conserve enstrophy (the integrated square of the fluid's vorticity field): this leads to an inverse energy cascade and a direct enstrophy cascade ~\cite{boffetta2012two}.
In 2D superfluids, vorticity is carried by singly charged quantized vortices, making enstrophy and vortex-number conservation equivalent.
Both numerical~\cite{numasato2010direct, numasato2010possibility,chesler2013holographic} and experimental~\cite{kwon2014relaxation,seo2017observation} studies indicate that vortex number is in general not conserved, and special care is required to avoid vortex recombination~\cite{bradley2012energy,neely2013characteristics,gauthier2019giant,johnstone2019evolution}.
Our numerics (see SM) as well as those of Ref.~\cite{numasato2010direct} show that  vortices are rapidly lost in our system leading to decaying Kolmogorov scaling with a direct energy cascade~\footnote{See the Supplementary Material for a discussion of the direction of the energy cascade, including Refs.~\cite{jun2005large,cerbus2013intermittency,daniel2000intermittency,benzi1993extended,paret1999vorticity,paret1998intermittency}.}.

A second question is why---both in experiment and numerics---does ${\rm PDF}(\Delta v^L)$ have an excess kurtosis that is independent of separation?
What is the relation between scaling observed from VSFs and that obtained from TOF momentum distributions~\cite{thompson2013evidence,navon2016emergence}? 
Additionally, our studies focused on decaying turbulence in which a turbulent state is in the process of relaxing;  this is in contrast with fully-developed (i.e., steady state) turbulence in which the system is continuously excited at long length scales and energy is removed at short scales.
The recent development of BECs undergoing continuous replenishment and evaporation~\cite{Chen2022} may enable access to this regime.

On the numerical side, the dGPE introduced dissipation in an experimentally motivated, but ultimately heuristic manner.  
Full 3D simulations including realistic modeling of the evaporation processes would eliminate the need for heuristics, and also inform more realistic approximate 2D descriptions.
Such simulations would help connect our results to those inferring turbulence from momentum distributions~\cite{navon2016emergence,karailiev2024observation}, that found wave-turbulence scaling rather than Kolmogorov scaling.

\begin{acknowledgments}
The authors thank E.~Mercado and Y.~Geng for carefully reading the manuscript.
This work was partially supported by the National Institute of Standards and Technology; the Quantum Leap Challenge Institute for Robust Quantum Simulation (OMA-2120757); and the Air Force Office of Scientific Research Multidisciplinary University Research Initiative ``RAPSYDY in Q'' (FA9550-22-1-0339).
\end{acknowledgments}

%

\end{document}